\documentclass[aps,showpacs,preprintnumbers,amsmath,amssymb]{revtex4}

\usepackage{float}
\usepackage{graphics,epsfig}
\usepackage{graphicx}
\usepackage{dcolumn}
\usepackage{bm}

\begin{document}
\baselineskip=0.8 cm
\title{{\bf Phantom scalar emission in the Kerr black hole spacetime}}
\author{Songbai Chen}
\email{csb3752@163.com}
\affiliation{Institute of Physics and
Department of Physics, Hunan Normal University,  Changsha, Hunan
410081, P. R. China \\ Key Laboratory of Low Dimensional Quantum
Structures \\ and Quantum Control of Ministry of Education, Hunan
Normal University, Changsha, Hunan 410081, P. R. China}

\author{Jiliang Jing}
\email{jljing@hunnu.edu.cn}
\affiliation{Institute of Physics and
Department of Physics, Hunan Normal University,  Changsha, Hunan
410081, P. R. China \\ Key Laboratory of Low Dimensional Quantum
Structures \\ and Quantum Control of Ministry of Education, Hunan
Normal University, Changsha, Hunan 410081, P. R. China}
\begin{abstract}
\baselineskip=0.6 cm
\begin{center}
{\bf Abstract}
\end{center}

We study the absorption probability and Hawking radiation spectra of
a phantom scalar field in the Kerr black hole spacetime. We find
that the presence of the negative kinetic energy terms modifies the
standard results in the greybody factor, super-radiance and Hawking
radiation. Comparing with the usual scalar particle, the phantom
scalar emission is enhanced in the black hole spacetime.

\end{abstract}

 \maketitle
\newpage

\section{Introduction}

Our Universe is assumed to be filled with dark energy because it can
explain the accelerated expansion of the Universe, which is strongly
supported by many cosmological observations\cite{1b,1b0}. Dark
energy is an exotic energy component with negative pressure and
constitutes about $72\%$ of present total cosmic energy. The leading
interpretation of such a dark energy is the cosmological constant
with an equation of state $\omega_x=-1$ (for a classic review see
\cite{1ax}, for a recent nice review see \cite{1ar0}, and for a
recent discussion see \cite{1ar1,1b1,1ar2}). The energy density of
this dark energy is associated with quantum vacuum
\cite{1ax,1ar0,1ar1,1ar2,1b1,1as}. Although this explanation is
consistent with observational data, it is plagued with the so-called
coincidence problem namely, ``why are the vacuum and matter energy
densities of precisely the same order today?". Therefore the
dynamical scalar fields, such as quintessence \cite{2a}, k-essence
\cite{3a} and phantom field \cite{4a}, are proposed as possible
alternatives of dark energy.

Comparing with other dynamical scalar fields, the phantom field
model is more interesting  because that it has a negative kinetic
energy and the super negative equation of state $\omega_x<-1$.
Although the null energy condition is violated, this dark energy
model is not ruled out by recent precise observational data
involving CMB, Hubble Space Telescope and type Ia Supernova
\cite{5a}. The dynamical evolution of the phantom field in the
cosmology has been investigated in the last years
\cite{6a1,6a11,6a2,6a3,6a4,6a5,6a6,6a7}. It shows that the energy
density increases with the time and approaches to infinity in a
finite time \cite{6a1}. This implies in the standard Einstein
cosmology the flat universe dominated by phantom energy will blow up
incessantly and arrive at a future singularity finally named big rip
which has such a strong exclusive force that anything in the
universe including the large galaxies will be torn up. Recently,
many efforts have been working to avoid the big rip \cite{7b}. It
has argued that this future singularity could be vanished in the
universe if one considers the effects from loop quantum gravity
\cite{lc3,lg4,lg6,lg7}.

The presence of negative kinetic energy results in many exotic
properties of phantom field in the black hole spacetime. E. Babichev
\textit{et al} \cite{Eph} considered the phantom energy accretion of
black hole and find that the mass of the black hole is decreased.
This can be explained by that the kinetic energy of the phantom
field is negative which yields the super negative equation of state
$\omega_x<-1$. The decrease of mass of black hole in the phantom
energy accretion will lead to the ratio between charge and mass of
black hole could be larger than $1$  ($\frac{Q}{M}>1$ ) and there
may exists a naked singularity \cite{Eph1}, which implies that the
cosmological censorship is violated. The negative kinetic energy
also yields that the dynamical evolution of phantom scalar
perturbation  possesses some unique characteristics in the black
hole spacetime \cite{sjp}. One of is that it grows with an
exponential rate in the late-time evolution rather than decays as
the usual scalar perturbations. These new results will excite more
efforts to be devoted to the study of phantom energy in the
background of a black hole. In this paper we will focus on the
Hawking radiation of the phantom scalar particles in the Kerr black
hole spacetime and see what effect of the negative kinetic energy on
the power and angular momentum emission spectra of the Hawking
radiation.

\section{Phantom scalar emission in the Kerr black hole spacetime}

In the curve spacetime, the action of the phantom scalar field with
the negative kinetic energy term is
\begin{eqnarray}
S=\int d^4x \sqrt{-g}\bigg[-\frac{R}{16\pi
G}-\frac{1}{2}\partial_{\mu}\psi\partial^{\mu}\psi+V(\psi)\bigg].
\end{eqnarray}
Here we take metric signature $(+ - - -)$ and the potential
$V(\psi)=-\frac{1}{2}\mu^2\psi^2$, where $\mu$ is the mass of the
scalar field. Varying the action with respect to $\psi$, we obtain
the Klein-Gordon equation for a phantom scalar field in the curve
spacetime
\begin{eqnarray}
\frac{1}{\sqrt{-g}}\partial_{\mu}(\sqrt{-g}g^{\mu\nu}\partial_{\nu})
\psi-\mu^2\psi=0.\label{WE}
\end{eqnarray}
The presence of negative kinetic energy leads to the sign of the
mass term $\mu^2$ is negative in the wave equation, which will yield
the peculiar properties of Hawking radiation of the phantom scalar
particle in the black hole spacetime.

The well-known Kerr metric in the Boyer-Lindquist coordinate is
\begin{eqnarray}
ds^2&=&\bigg(1-\frac{2Mr}{\Sigma}\bigg)dt^2+\frac{4Mra\sin^2{\theta}}{\Sigma}dtd\phi-\frac{\Sigma}{\Delta}dr^2
-\Sigma
d\theta^2-\bigg(r^2+a^2+\frac{2Mra^2\sin^2{\theta}}{\Sigma}\bigg)\sin^2{\theta}d\phi^2,\label{m1}
\end{eqnarray}
with
\begin{eqnarray}
\Delta=r^2-2Mr+a^2,\;\;\;\;\;\;\;\;\;\;\;\;\Sigma=r^2+a^2\cos^2{\theta},
\end{eqnarray}
where $M$ is the mass and $a$ is the angular momentum of the black
hole. Equation (\ref{WE}) is separable in terms of the spheroidal
harmonics $\psi(t,r,\theta,\phi)=e^{i(-\omega
t+m\phi)}R(r)S(\theta)$. The angular and the radial functions
$S(\theta)$, $R(r)$ obey to
\begin{eqnarray}
\frac{1}{\sin{\theta}}\frac{d}{d\theta}\bigg[\sin{\theta} \frac{d
S(\theta)}{d \theta}\bigg]
+\bigg[(\omega^2+\mu^2)a^2\cos^2{\theta}-\frac{m^2}{\sin^2{\theta}}+\lambda\bigg]S(\theta)=0,\label{angd}
\end{eqnarray}
and
\begin{eqnarray}
\frac{d}{dr}\bigg[\Delta\frac{d R(r)}{dr}\bigg]
+\bigg\{\frac{[(r^2+a^2)\omega-ma]^2}{\Delta}+\mu^2r^2-E_{lm}\bigg\}R(r)=0,\label{radial}
\end{eqnarray}
respectively. Where $\lambda$ is the eigenvalues and the function
$E_{lm}=\lambda+\omega^2a^2-2ma\omega$. In order to calculate the
absorption probability $|\mathcal{A}_{lm}|^2$ and the luminosity of
Hawking radiation for a phantom scalar particle, we must solve the
radial equation (\ref{radial}) above. Following the standard
matching techniques
\cite{Haw3,Kanti,Kan1,Kan2,Kan3,Kan4,Kan5,Kan6,Haw4,Haw5}, we can
create a smooth analytical solution of the equation (\ref{radial})
in the low-energy and low-angular momentum limit. Near the horizon
($r\simeq r_+$) regime and at infinity, it has the form
\begin{eqnarray}
&&r\rightarrow r_+,~~~~~~
R(r=r_+)=A^{(r=r_+)}_{in}e^{-(\omega-m\Omega_H)r_*}+A^{(r=r_+)}_{out}e^{(\omega-m\Omega_H)r_*},\\
&&r\rightarrow \infty,~~~~~~
R(r=\infty)=A^{(r=\infty)}_{in}e^{-\sqrt{(\omega^2+\mu^2)r}}+A^{(r=\infty)}_{out}e^{\sqrt{(\omega^2+\mu^2)r}},
\end{eqnarray}
respectively. Unlike the usual scalar particle, we find that for the
phantom particle with an arbitrary value of $\mu$ the solution above
denotes an incoming and an outgoing spherical waves at large
distance from the black hole. From this solution, we can calculate
the absorption probability
\begin{eqnarray}
|\mathcal{A}_{lm}|^2=1-\bigg|\frac{A^{(r=\infty)}_{out}}{A^{(r=\infty)}_{in}}\bigg|^2.\label{GFA}
\end{eqnarray}
The power and and angular momentum emission spectra of phantom
scalar particle is then written as
\begin{eqnarray}
\frac{d^2E}{dtd\omega}&=&\sum_{l,m}
\frac{\omega^2}{e^{\;(\omega-m\Omega_H)/T_{H}}-1}\frac{N_l|\mathcal{A}_{lm}|^2}{2\pi\sqrt{\omega^2+\mu^2}},\label{H1}\\
\frac{d^2J}{dtd\omega}&=&\sum_{l,m}
\frac{m\omega}{e^{\;(\omega-m\Omega_H)/T_{H}}-1}\frac{N_l|\mathcal{A}_{lm}|^2}{2\pi\sqrt{\omega^2+\mu^2}},
\end{eqnarray}
where $T_H$ is the Hawking temperature of Kerr black hole. These
equations can be integrated numerically.

Here we present the numerical results about the absorption
probability $|\mathcal{A}_{lm}|^2$ and the Hawking radiation of a
phantom scalar field in the background of a Kerr black hole.

In Fig.(1), we fix $a=0.2$ and examine the dependence of the
absorption probability of phantom scalar particle on its mass $\mu$
for the first partial waves ($l=0, \;m=0$) in the background of a
Kerr black hole. Comparing with the usual scalar particle, the
absorption probability of the phantom increase rather than decrease
as the mass $\mu$ increases. This is not surprising because that in
the wave equation of phantom field (\ref{WE}) the sign of the mass
term is negative get rise to the absorption probability has the form
$|\mathcal{A}_{00}|^2\sim 4\omega \sqrt{\omega^2+\mu^2}(r^2_+-a^2)$
in the low-energy and low-angular momentum limit. For other values
of $l$, we also find that the absorption probability of phantom
scalar particle increase with the mass $\mu$.

\begin{figure}[ht]
\begin{center}
\includegraphics[width=8.0cm]{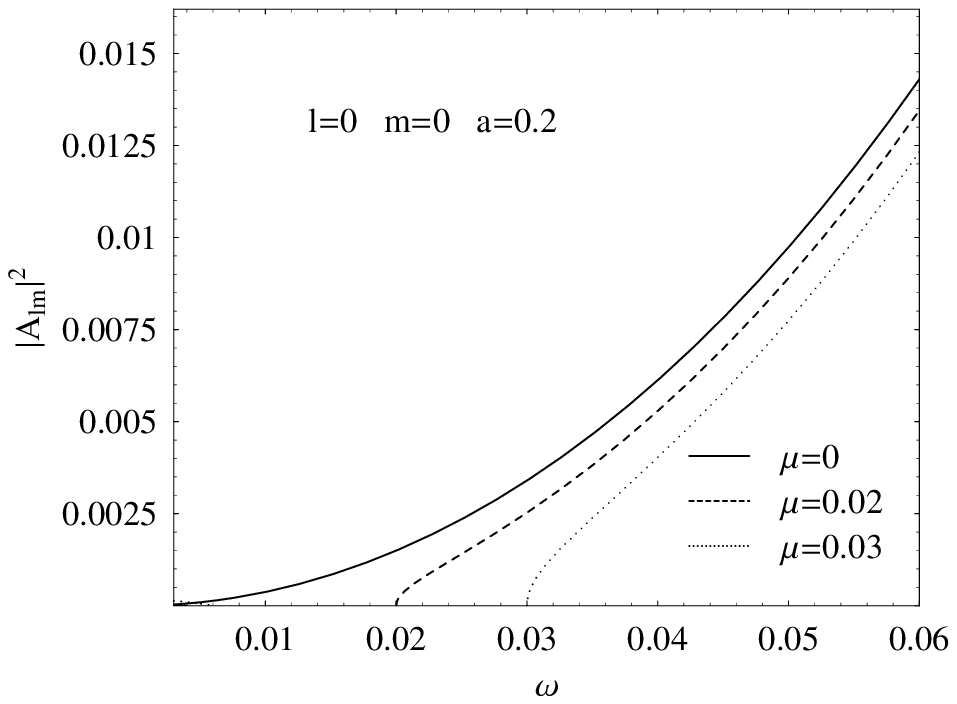}\;\;\;\;\includegraphics[width=8.0cm]{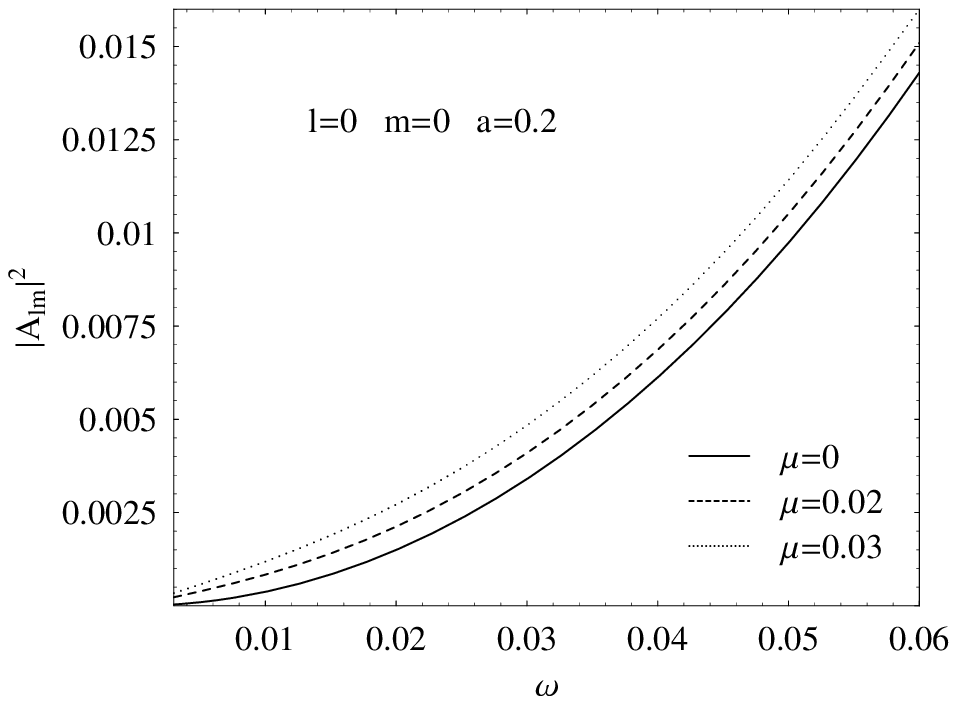}\;\;\;\;\;
\caption{Variety of the absorption probability
$|\mathcal{A}_{lm}|^2$ of a
 scalar particle propagating in the Kerr black hole with its mass $\mu$ (the left panel is for a usual scalar particle
 and the right panel is for a phantom one) for fixed $l=0$, $m=0$ and $a=0.2$. We set $M=1$.}
 \end{center}
 \label{fig1}
\end{figure}

\begin{figure}[ht]
\begin{center}
\includegraphics[width=8.0cm]{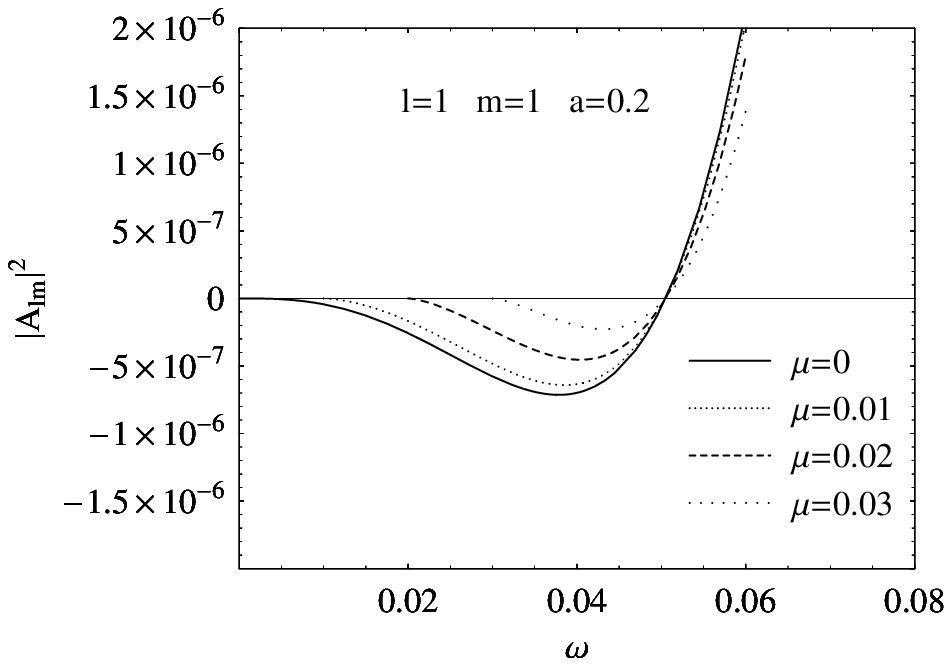}\;\;\;\;\;\;\includegraphics[width=8.0cm]{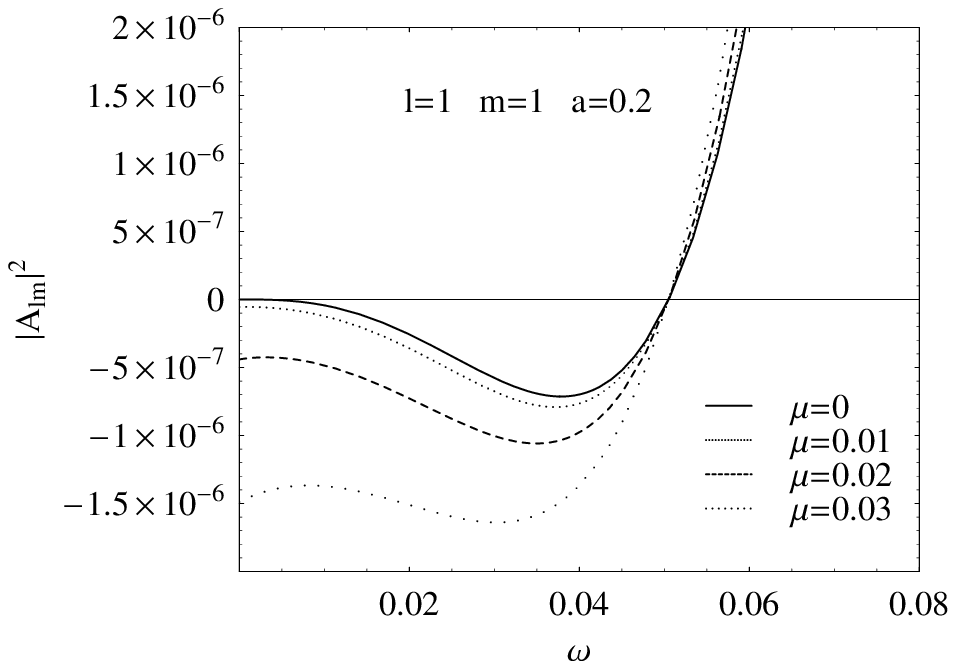}
\caption{The dependence of the super-radiance on the mass $\mu$ (the
left panel is for a usual scalar particle and the right panel is for
a phantom one) in the Kerr black hole for fixed $l=1$, $m=1$ and
$a=0.2$. We set $M=1$.}
\end{center}
\label{fig2}
\end{figure}
\begin{figure}[ht]
\begin{center}
\includegraphics[width=8.0cm]{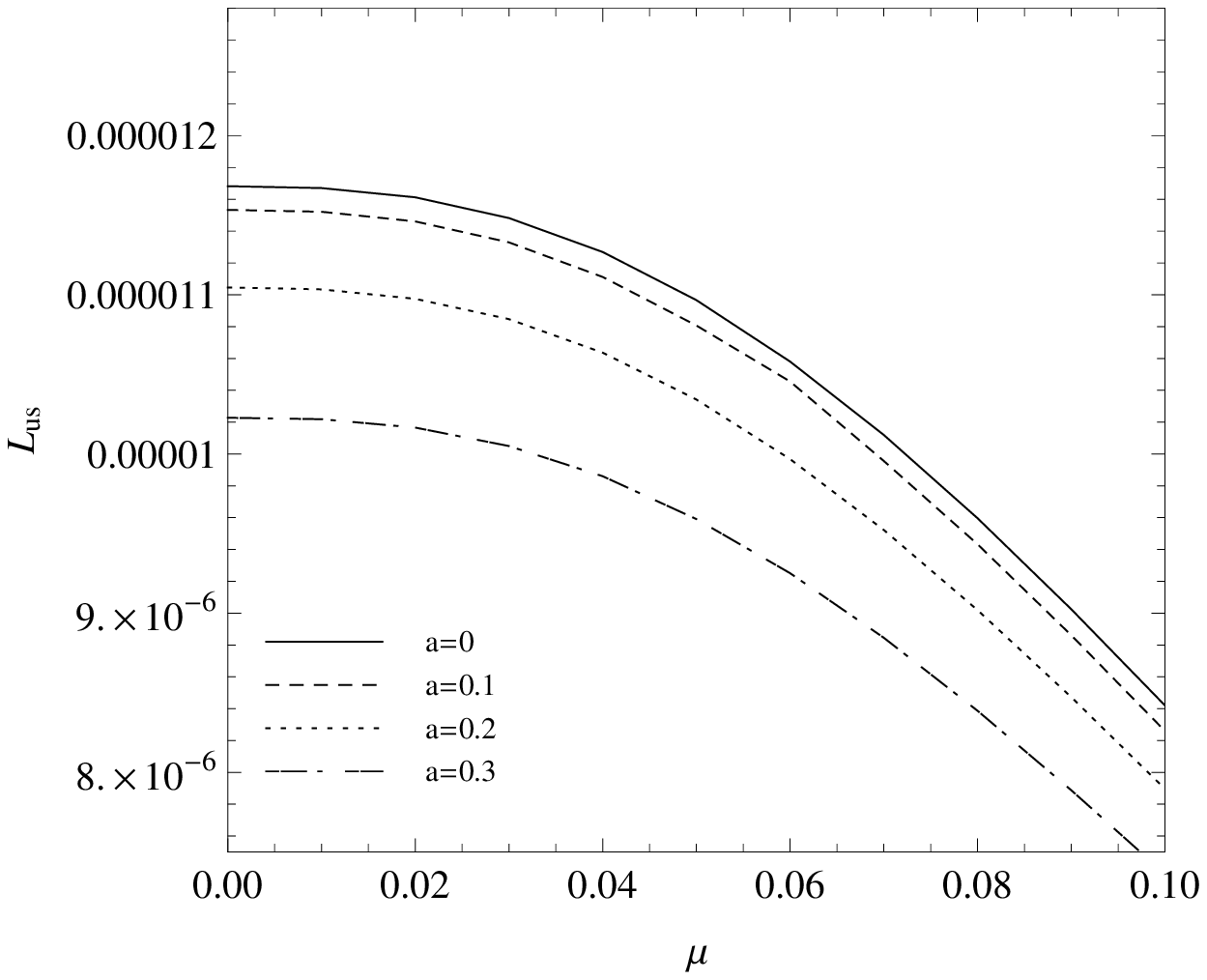}\;\;\;\;\; \includegraphics[width=8.0cm]{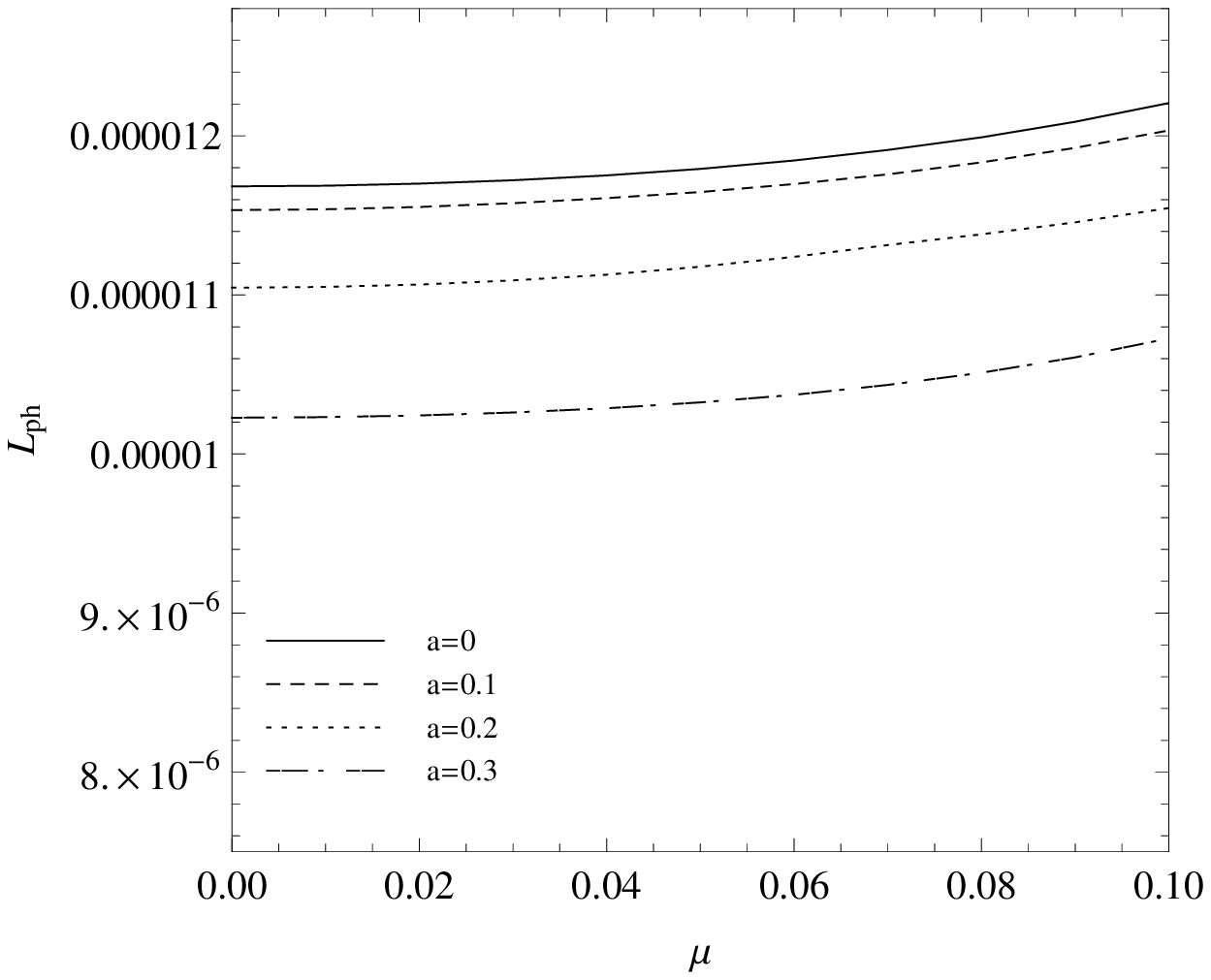}
\caption{Variety of the luminosity of Hawking radiation $L$ of
scalar particles propagating with $\mu$ in the Kerr black hole, for
$l=0, m=0$ and different $a$. We set $M=1$. The left panel for a
usual scalar particle and the right panel for a phantom particle.}
\end{center}
\label{fig3}
\end{figure}
\begin{figure}[ht]
\begin{center}
\includegraphics[width=8.0cm]{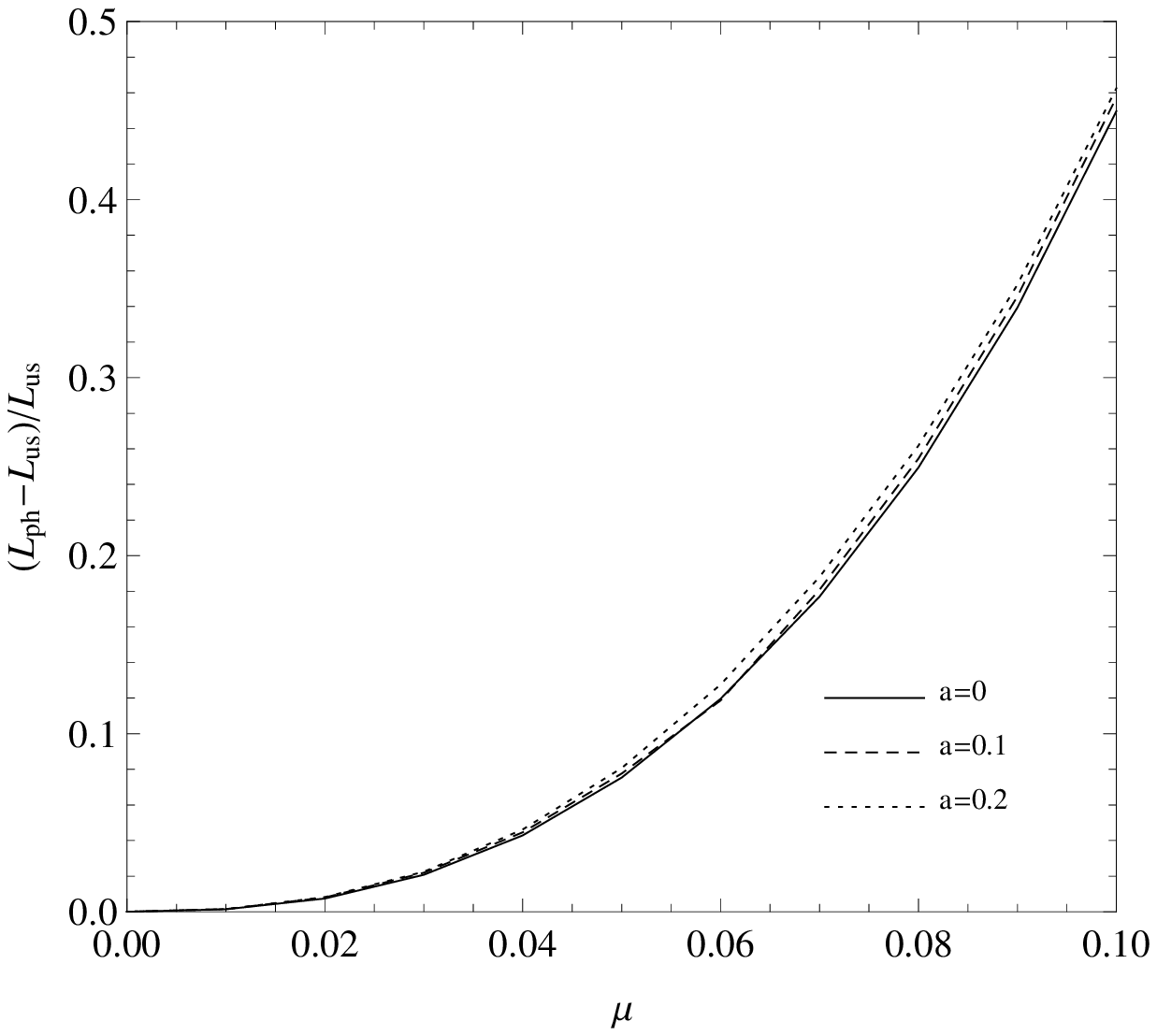}
\caption{The relative luminosity of Hawking radiation between
phantom and usual massive scalar particles (with the same mass as
the phantom field) for different values of $\mu$ in the Kerr black
hole. We set $M=1$.}
\end{center}
\label{fig4}
\end{figure}

In Fig.(2), we plot the absorption probability of phantom particle
for the mode ($l=1,m=1$) and find that the super-radiance possesses
some peculiar properties in this case. First, for phantom particle
the range of $\omega$ for the super-radiance to happen is
independent of $\mu$, which means that the phantom scalar particle
can be radiated away without any constraint of the particle's
energy. This can be explained by that the wave-function of phantom
particle at the spatial infinity has the form
$e^{i\sqrt{\omega^2+\mu^2}\;r_*}$. It is different from that of the
usual scalar particles. It is well known that the usual scalar
particle with the energy $\omega^2<\mu^2$ cannot be radiated away.
These particles are is forced to be trapped in the system and the
amplified wave can accumulate in the potential well, which will
yield black hole bomb in which the mass $\mu$ of a usual scalar
particle plays the role of a natural mirror. While in the
super-radiance of the phantom scalar particle, it can not yield
black hole bomb since all of phantom scalar particles can be
radiated away and their wave functions are not in the bounded-state.
Second, the magnitude of the super-radiance of phantom particle
increases with the mass $\mu$. While for usual scalar particle, it
is decreases. The new properties of super-radiance imply that the
phantom field will enhance Hawking radiation of the black hole.

\begin{figure}[ht]
\begin{center}
\includegraphics[width=8.0cm]{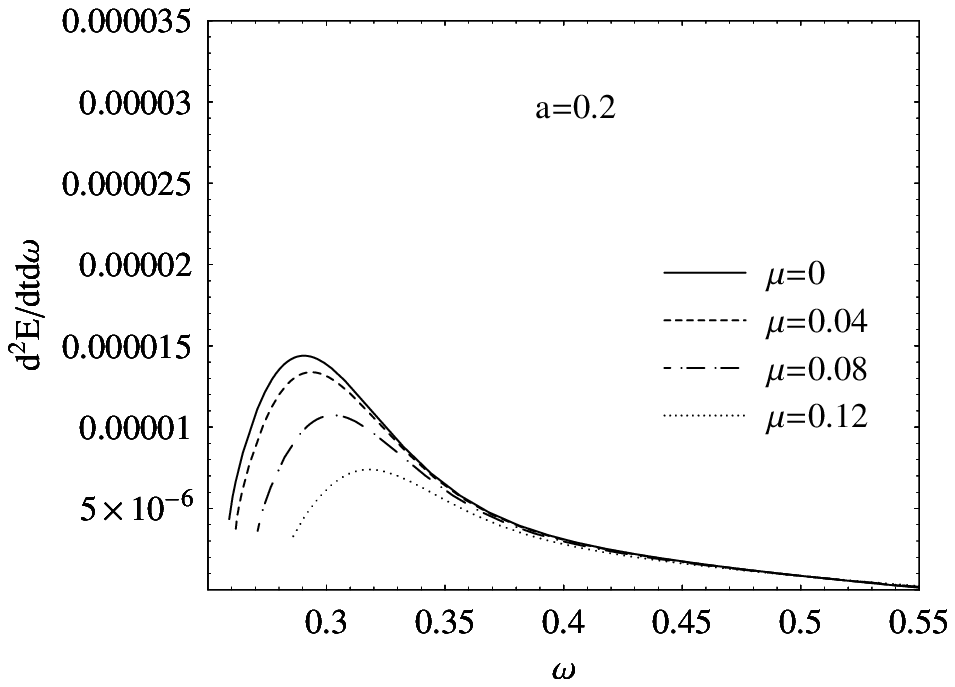}\;\;\;\;\; \includegraphics[width=8.0cm]{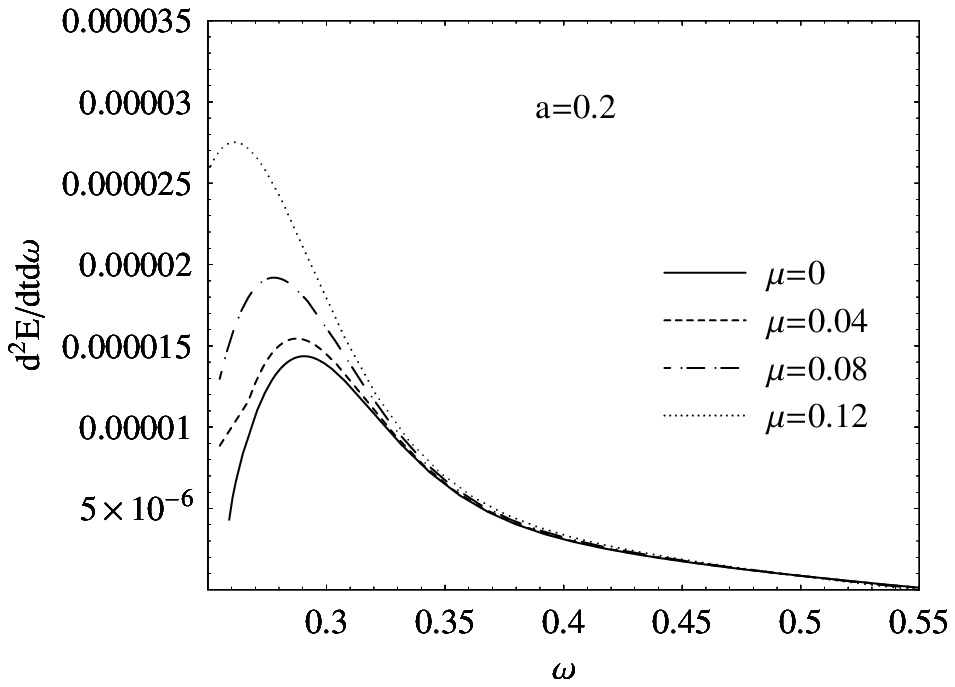}
\caption{Power emission spectra  of usual scalar particle (the left
panel) and of the phantom scalar one (the right panel) for different
values of $\mu$ in the Kerr black hole. We set $M=1$.}
\end{center}
\label{fig5}
\end{figure}

\begin{figure}[ht]
\begin{center}
\includegraphics[width=8.0cm]{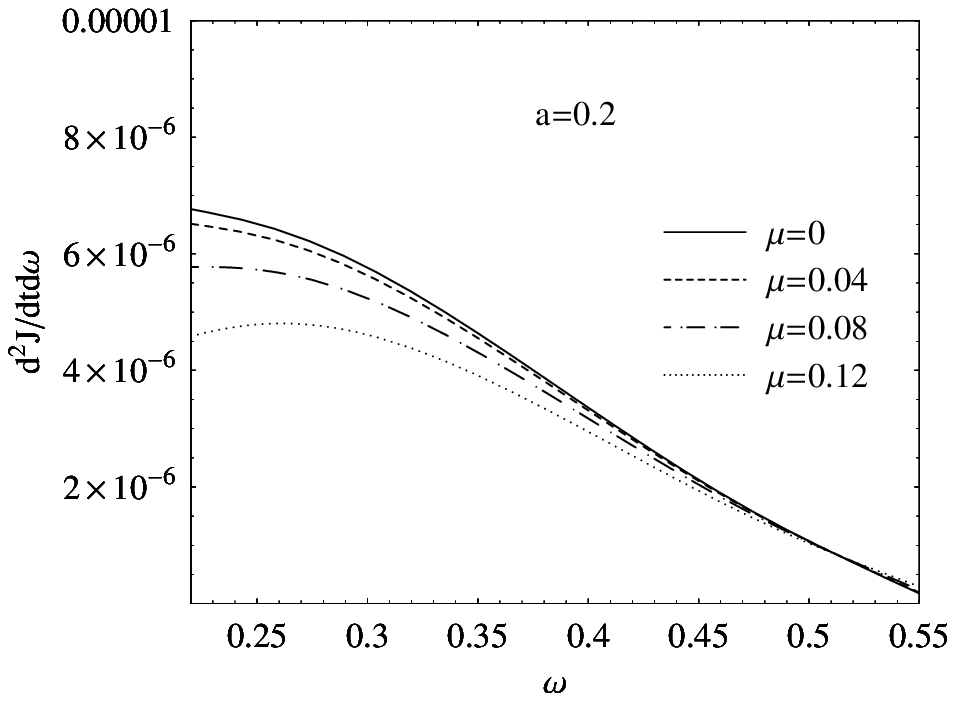}\;\;\;\;\; \includegraphics[width=8.0cm]{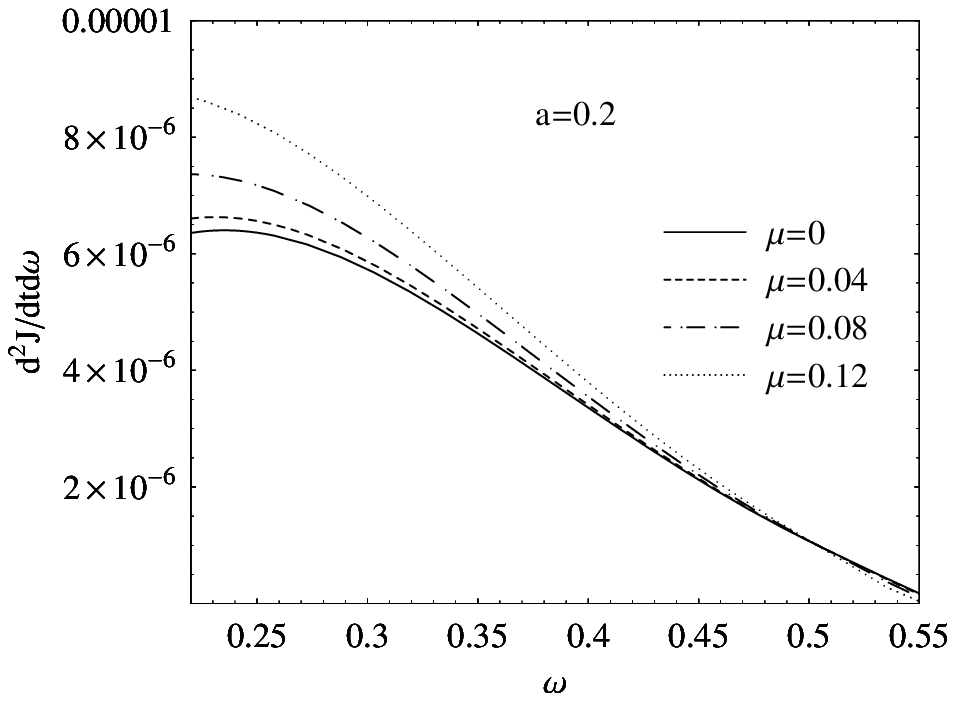}
\caption{Angular momentum spectra of usual scalar particle (the left
panel) and of the phantom scalar one (the right panel) for different
values of $\mu$ in the Kerr black hole. We set $M=1$.}
\end{center}
\label{fig6}
\end{figure}

Let us now to study the luminosity of the Hawking radiation of
phantom particles in the Kerr black hole spacetimes. We first focus
on the first partial waves ($l=0$, $m=0$), which plays a dominant
role in the greybody factor. The luminosity of the Hawking radiation
for the phantom scalar particle can be expressed as
\begin{eqnarray}
L&=&\int^{\infty}_0
\frac{|\mathcal{A}_{00}|^2}{\sqrt{\omega^2+\mu^2}}\frac{\omega^2}{e^{\;(\omega-m\Omega_H)/T_{H}}-1}\frac{d\omega}{2\pi}.
\label{LHK}
\end{eqnarray}
Here the low bound of the integral is zero rather than $\mu$ since
the emission of the phantom scalar particle is not constrained by
the value of the particle mass $\mu$. In generally, this integral
cannot be computed analytically. In figure (3), we present some
numerical results. For the fixed $\mu$, the Hawking radiation of the
phantom scalar particles decreases as the angular momentum a
increases. This is the same as that of the usual scalar particles in
the Kerr black hole. But for the fixed $a$, with the mass $\mu$
increase the Hawking radiation of black hole increases for the
phantom scalar field  and  decreases for the usual scalar field.
Therefore the presence of negative kinetic leads to that the
emission of phantom field increases and black hole evaporates more
quickly. Figure (4) also tell us that with the increase of the mass
$\mu$ the difference between the emission of phantom and usual
scalar particle increase rapidly. When $\mu=0.1$, the luminosity of
Hawking radiation of phantom field is almost 1.5 multiples of that
of the usual scalar one for all $a$. Thus for the larger $\mu$,  the
Hawking radiation of black hole is dominated by phantom scalar
field. In Figs.(5) and (6), we summed up to $l=4$ modes in
calculating the energy and angular momentum emission rates. We find
that the mass $\mu$ of phantom scalar particle enhance the the power
and angular momentum emission spectra. Our results are consistent
with that obtained in the phantom energy accretion of black hole
where the black hole mass is decreased. This leads to increase of
the temperature of the black hole and enhance the luminosity of the
Hawking radiation.

\section{summary}

In this paper, we have studied the greybody factor and Hawking
radiation for a phantom scalar field in the background of Kerr black
hole in the low-energy and low-angular momentum approximation. We
have found that the absorption probability and Hawking radiation
contain the imprint of the phantom scalar field. The effects of
$\mu$ on the super-radiance and the luminosity of the Hawking
radiation for phantom scalar field are different from that for the
usual scalar field since phantom scalar particle possesses a
negative kinetic energy. Our results may help us to detect whether
our universe is filled with phantom field or not. It would be of
interest to generalize our study to other black hole spacetimes,
such as stringy black holes etc. Work in this direction will be
reported in the future.

\begin{acknowledgments}

This work was partially supported by the National Natural Science
Foundation of China under Grant No.10875041 and the construct
program of key disciplines in Hunan Province. J. L. Jing's work was
partially supported by the National Natural Science Foundation of
China under Grant No.10675045, No.10875040 and No.10935013; 973
Program Grant No. 2010CB833004 and the Hunan Provincial Natural
Science Foundation of China under Grant No.08JJ3010.
\end{acknowledgments}

\end{document}